# Theoretical and Experimental Study on Heat Transfer Characteristics of Water Heat Pipe

+Ziyi Wang[1], Huang Zhang[2,*], Shanfang Huang[1]

1. Department of Engineering Physics, Tsinghua University, Beijing 100084, China

2. Institute of Nuclear and New Energy Technology, Key Laboratory of Advanc0ed Reactor Engineering and Safety, Tsinghua University, Beijing 100084, China

Corresponding author: zhanghuang@tsinghua.edu.cn

**Abstract:** Heat pipe is an efficient heat transfer element based on two-phase natural circulation, which has advantages of simple structure, strong heat transfer ability, and good isothermal performance. Heat pipes are widely used in heat transfer and other fields, and especially have important applications in nuclear engineering. One of its most important characteristics is the heat transfer limit. In this work, heat transfer limits are first reviewed, and the detailed calculation equations are presented. Then, a Matlab code to calculate heat transfer limits as well as the thermal conductance are provided. Second, an experimental setup for testing the heat transfer characteristics of heat pipes was developed, which could be used to measure the thermal conductance as well as the heat transfer limits. The calculated results show that, for water heat pipes, the capillary limit mainly affects heat transfer in low temperature conditions, while in high temperature conditions, boiling limit dominates. In addition, the experiment results show that the thermal conductance of the measured heat pipe is 7267 W/(m$^2$·K), which agrees with the calculation result.

# 1. Introduction

Since Grover invented the heat pipe in 1964[1], it has been widely used in thermal engineering, aerospace engineering, and civil engineering. Recently, the success of Kilowatt Reactor Using Stirling TechnologY (KRUSTY) shows that heat pipes have great potential applications in nuclear engineering[2].

Figure 1 shows the structure and working principle inside a simple heat pipe. A heat pipe has a vacuum sealed metal shell, and the inner surface of the shell is covered with a capillary wick, which is filled with working fluid. When the heat pipe is in operation, the working fluid in the capillary wick of the evaporator section is heated and evaporated. The gas-phase working fluid flows along the vapor channel towards the condenser section under the pressure drop of vapor, and gradually releases heat and condenses in the condenser section before entering the capillary wick again. The capillary wick then transports the liquid in the condenser section back to the evaporation section through capillary force, completing a cycle of heat transfer[3].

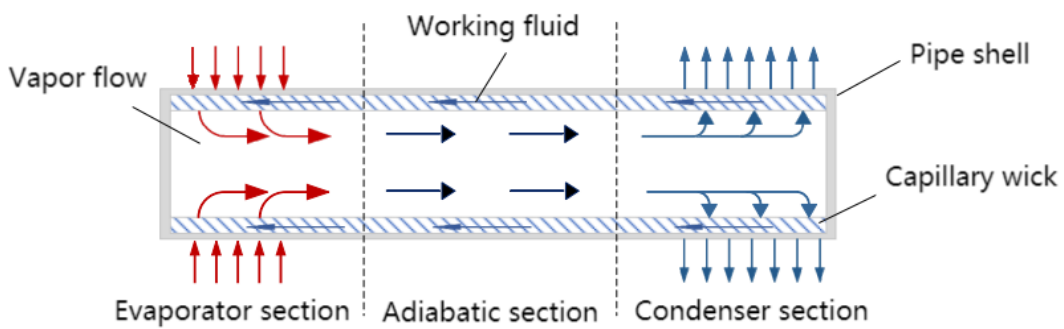

**Figure 1** Structure and working principle of heat pipe

Although heat pipes can transport a very large amount of heat, the heat flux of each heat pipe is ultimately limited. Once the heat flux or heat low exceeds the limit that the heat pipe can withstand, the thermal conductivity of the heat pipe will rapidly decrease. The limit of maximum heat flux or maximum heat flow of a heat pipe is called the "heat transfer limit of the heat pipe", abbreviated as the "limit of the heat pipe". There are many situations where heat pipes reach their limits, the influence factors include vapor flow velocity, vapor pressure, capillary force of the capillary wick, boiling of the fluid, friction between the gas and liquid phases of the working fluid, and so on. These heat transfer limits are related to the size of heat pipes, the structure of the capillary wick, working conditions, and the type of working fluid, as shown in Figure 2.

This work introduces a one-dimensional model for heat transfer limit of heat pipes, which includes sonic limits, viscous limit, capillary limit, entrainment limit and boiling limit, and an analysis and calculation program for the heat transfer limit of heat pipes was developed. Second, a thermal conductance calculation method for heat pipes was established. Third, an experimental platform for testing the water heat transfer characteristics of heat pipes was developed, and their thermal conductance were measured. Finally, the calculation and experimental results were compared.

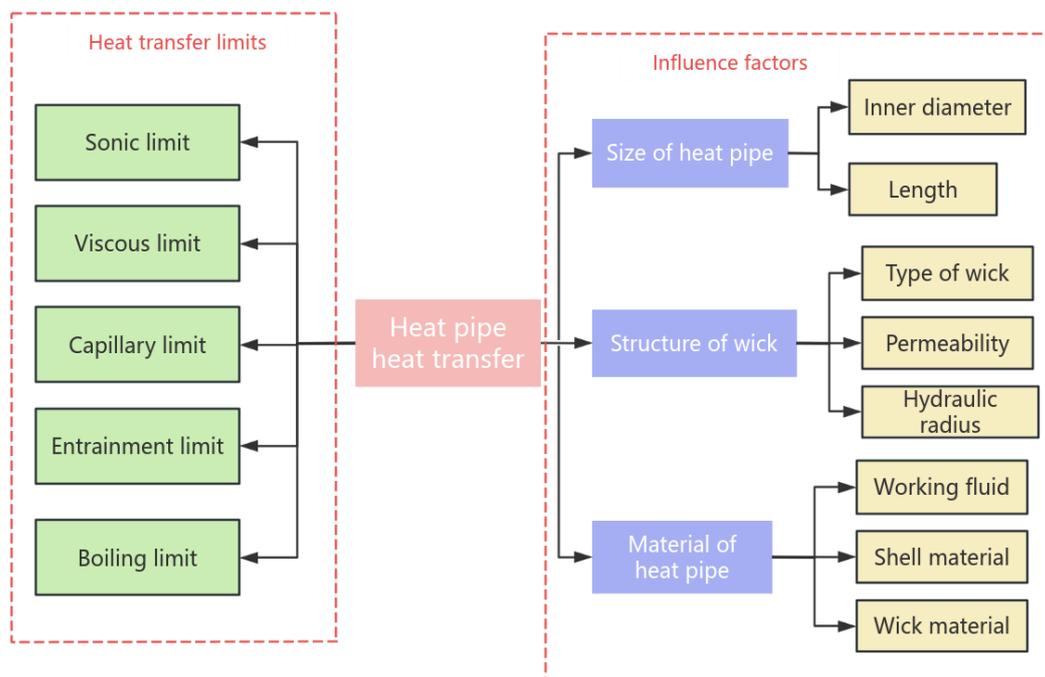

**Figure 2** Factors affecting heat transfer characteristics of heat pipes

## 2. Materials and Methods

**2.1 Theoretical Models**

2.1.1 Heat Transfer Limits

This section is going to introduce some typical models of heat transfer limit of heat pipe. And the derivation process of these models could be seen in the Appendix.

Assumptions are shown as: (1) the vapor is ideal gas, (2) the frictional effects may be neglected, (3) the inertial effects dominate, Busse derived the equation of sonic heat transfer limit of heat pipes as follows[4]:

$$Q_s = 0.474 A_v \lambda (\rho_0 P_0)^{\frac{1}{2}} \tag{1}$$

where both $\rho_0$ and $P_0$ are vapor density and pressure at the evaporator exit of heat pipe, $A_v$ is the vapor cross section area and $\lambda$ the latent heat of vaporization. Once the axial heat flux of the heat pipe reaches $q_s = Q_s/A_v$, the vapor flow is limited by the choking phenomenon. The axial heat flux then can no longer be increased by a decrease of the pressure in the condenser section(but only by an increase of the pressure and hence the temperature in the evaporator section). In choked inertia flow the vapor leaves the evaporator with sonic speed.

As the length of the heat pipe increases, the friction also increases, and the frictional effects on vapor flow gradually exceed that of inertial effects. When the pressure drop reaches zero at a point in the heat pipe due to friction, the vapor velocity no longer increases, resulting in the inability to further increase the heat flow and reach the heat transfer limit, which is called the viscous limit (or vapor-pressure limit)[2]:

$$Q_v = \frac{A_v r_v^2 \lambda}{16 \mu_v l_{\text{eff}}} \rho_0 P_0 \tag{2}$$

where $r_v$ is the radius of open heat pipe allowing vapor to pass through, $\mu_v$ the vapor viscosity in the evaporator and $l_{\text{eff}}$ the effective length of heat pipe.

Since heat pipes rely on capillary force to transport fluid, the total pressure used for transporting fluids in the wick should not exceed the sum of the maximum axial capillary pressure and the axial component of gravity. Then the equilibrium equation between maximum capillary pressure and fluid transport pressure in heat pipes should be

$$P_t = \frac{2\sigma}{r_c} - \rho_l g d \cos\varphi \tag{3}$$

where $\sigma$ is the surface tension coefficient, $r_c$ the effective capillary radius, $\rho_l$ the liquid density, g the gravitational acceleration, $d$ the internal diameter, and $\varphi$ the heat pipe inclination. By solving the equation above, Chi obtain the capillary heat transfer limit as follows[5]:

$$Q_c = \frac{(\frac{2\sigma}{r_c} - \rho_l g d \cos\varphi - \rho_l g l \sin\varphi)}{(F_v + F_l)(0.5 l_e + l_a + 0.5 l_c)} \tag{4}$$

where $l$ is the length of heat pipe, both $F_v$ and $F_l$ are vapor frictional coefficient and liquid frictional coefficient, $l_e$ the length of evaporator section, $l_a$ the length of adiabatic section and $l_c$ the length of condenser section.

It is known that in the cycle inside the heat pipe, the vapor flow moves from the evaporator section to the condenser section to transport heat, while the liquid returns from the condenser section to the evaporator section. Therefore, due to the opposite flow direction of the two, there will be shear forces at the gas-liquid interface. As the heat flow of the heat pipe increases and the vapor flow velocity continues to rise, the liquid may be entrained by the vapor flow from the capillary wick, then the heat transfer can be limited. Once the entrainment occurs, the flow of the fluid cycle will be significantly decreased, causing the wick in the evaporator section to burn out. The entrainment limit is given by[6]

$$Q_e = A_v \lambda \sqrt{\frac{\sigma \rho_l}{2 r_h}} \tag{5}$$

where $r_h$ is the hydraulic radius of the wick surface pores.

A typical cylindrical heat pipe receives heat at the evaporator end where it is transferred to the working fluid radially. When the input flux is sufficient, nucleation sites are formed inside the wick and bubbles are trapped in the wick, blocking liquid return that results in evaporator dryout. As compared to other heat pipe limits, boiling limit is a radial flux constraint and not an axial flux constraint. The maximum heat flow beyond which bubble growth will occur resulting in dryout is given by[7]

$$Q_b = \frac{2\pi l_e k_\varepsilon T_v}{\lambda \rho_v \ln(r_i/r_v)} \left(\frac{2\sigma}{r_b}\right) \tag{6}$$

where $r_b$ is nucleation site radius (assume to be $2.54 \times 10^{-7}$ m), $T_v$ vapor temperature, $r_i$ inner radius of heat pipe, and $k_\varepsilon$ thermal conductivity coefficient of capillary wick.

2.1.2 Thermal Conductance

When a heat pipe operates under suitable conditions, its characteristic of heat transfer can be described by the thermal conductance. Given the heat flow $Q$, the definition of the thermal conductance $k_{\text{HP}}$ is as follows:

$$Q = A k_{\text{HP}} (T_e - T_c) \tag{7}$$

where both $T_e$ and $T_c$ are temperatures of evaporator section and condenser section.

As shown in Figure 3, heat transfer in heat pipes can be divided into five parts based on the cycle: heat is input from the shell of the evaporator section, absorbed by the fluid through the capillary wick of the evaporator section, transported to the condenser section through the vapor flow, condensed and released by the fluid through the wick of the condenser section, and finally outputted from the shell of the condensation section.

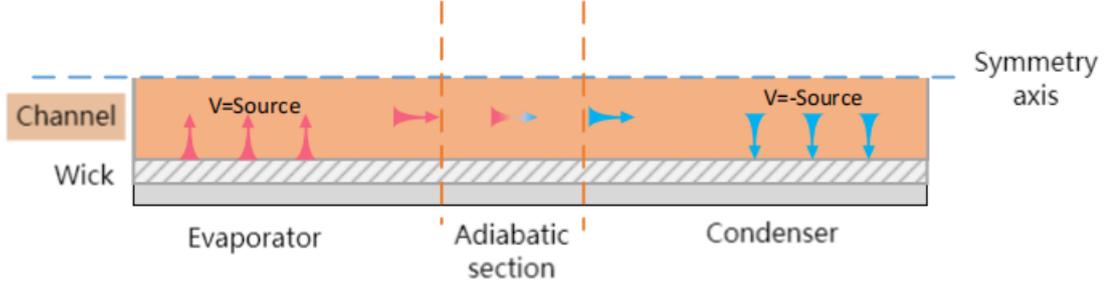

**Figure 3** The heat transfer process inside the heat pipe

The thermal conduction in the shell and wick of heat pipe can be described by the Fourier's low, and the heat transfer by vapor flow can be calculated based on the Clausius-Clapeyron equation. Moreover, the temperature difference at the gas-liquid interface is small and can be ignored for heat transfer. Applying the above equation to various parts of the heat flow path connected in series inside the pipe, it goes that:

shell thermal conduction in evaporator section:

$$T_{o,e} - T_{i,e} = \frac{\ln(r_o/r_i)}{2\pi l_e k_p} Q \quad (8)$$

wick thermal conduction in evaporator section:

$$T_{i,e} - T_{w,e} = \frac{\ln(r_i/r_v)}{2\pi l_e k_\varepsilon} Q \quad (9)$$

vapor flow thermal conduction:

$$T_{v,e} - T_{v,c} = \frac{T_v(P_{v,e} - P_{v,c})}{\rho \lambda} \quad (10)$$

wick thermal conduction in condenser section:

$$T_{w,c} - T_{i,c} = \frac{\ln(r_i/r_v)}{2\pi l_c k_\varepsilon} Q \quad (11)$$

shell thermal conduction in evaporator section:

$$T_{i,c} - T_{o,c} = \frac{\ln(r_o/r_i)}{2\pi l_c k_p} Q \quad (12)$$

by adding the above five equations and approximating them, with the definition of $k_{HP}$ given by Equation (7), the equation for the thermal conductance is derived as follows:

$$k_{HP} = \left[\frac{r_o t_p}{2l_e k_p} + \frac{r_o^2 t_w}{2l_e r_i k_\varepsilon} + \frac{\pi r_o^2 T_v(P_{v,e} - P_{v,c})}{\rho \lambda Q} + \frac{r_o^2 t_w}{2l_c r_i k_\varepsilon} + \frac{r_o t_p}{2l_c k_p}\right]^{-1} \quad (13)$$

where $(P_{v,e} - P_{v,c})$ is the sum of the vapor pressure drop across the entire section of the heat pipe. Assuming that the heat flux along the direction of vapor flow is consistent and uniform, the vapor pressure drop can be integrated to obtain:

$$(P_{v,e} - P_{v,c}) = F_v Q(\frac{l_e}{6} + l_a + \frac{l_c}{6}) \tag{14}$$

by substituting the above equation into Equation (13), calculation equation for $k_{HP}$ is as follows:

$$k_{HP} = \frac{1}{\frac{r_o t_p}{2l_e k_p} + \frac{r_o^2 t_w}{2l_e r_i k_\varepsilon} + \frac{\pi r_o^2 F_v(\frac{l_e}{6} + l_a + \frac{l_c}{6})T_v}{\rho \lambda} + \frac{r_o^2 t_w}{2l_c r_i k_\varepsilon} + \frac{r_o t_p}{2l_c k_p}}$$

$$= \frac{1}{R_{p,e} + R_{w,e} + R_v + R_{w,c} + R_{p,c}} \tag{15}$$

The thermal conductivity coefficient of the pipe shell $k_p$ is easy to know. $k_\varepsilon$ of some types of wick are collected in Table 1 from Chi[5].

Table 1 Expressions of $k_\varepsilon$ for liquid-saturated wicks[5]

| Wick structure | $k_\varepsilon$ Expressions |
| --- | --- |
| Rectangular grooves | $k_\varepsilon = \dfrac{\omega_f k_l k_w \delta + w k_l(0.185\omega_f k_w + \delta k_l)}{(\omega + \omega_f)(0.185\omega_f k_w + \delta k_l)}$ |
| Wire screen | $k_\varepsilon = \dfrac{k_l[(k_l + k_w) - (1 - \varepsilon)(k_l - k_w)]}{[(k_l + k_w) + (1 - \varepsilon)(k_l - k_w)]}$ |
| Packed sphere | $k_\varepsilon = \dfrac{k_l[(2k_l + k_w) - 2(1 - \varepsilon)(k_l - k_w)]}{[(2k_l + k_w) + (1 - \varepsilon)(k_l - k_w)]}$ |

$k_w$ and $k_l$ represent the coefficient of thermal conductivity of the wick material and the working fluid; $\omega_f$, $\omega$ and $\delta$ are the groove fin thickness, groove thickness, and groove depth, respectively; $\varepsilon$ is the porosity of the wick.

**2.2 Experimental Setup**

Due to the complex physical processes inside heat pipes, many characteristics used for heat pipe calculations are obtained through experiments，and the above equation also uses a simplified heat transfer model for heat pipes in the derivation. Therefore, we need to conduct an experimental setup to verify the heat transfer effects of the heat pipe. Figure 4 shows the schematic diagram of the experimental setup used in this work. This setup is divided into four parts: heat pipes and brackets, heating system, cooling system, and measurement system. Detailed information of this setup is shown in Figure 5.

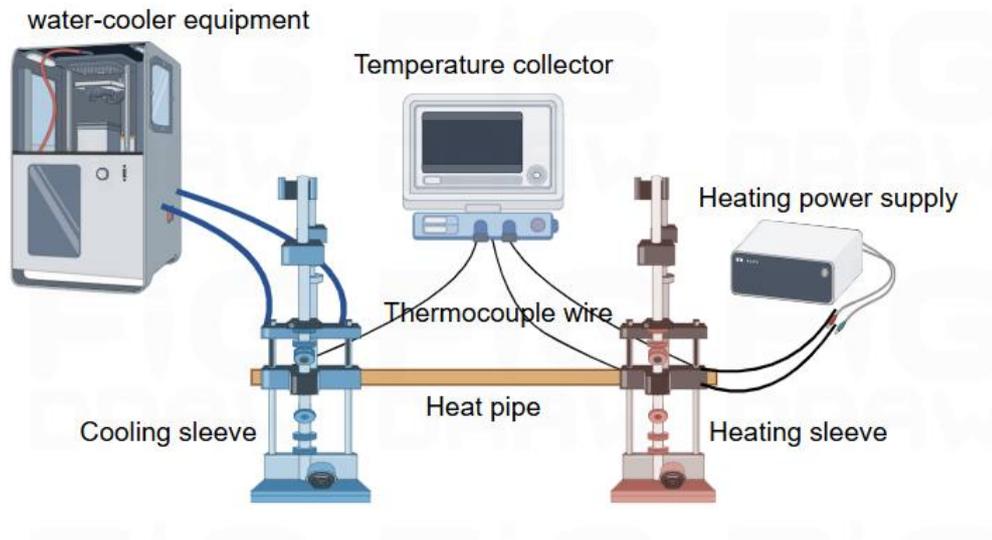

**Figure 4** Schematic diagram of experimental system

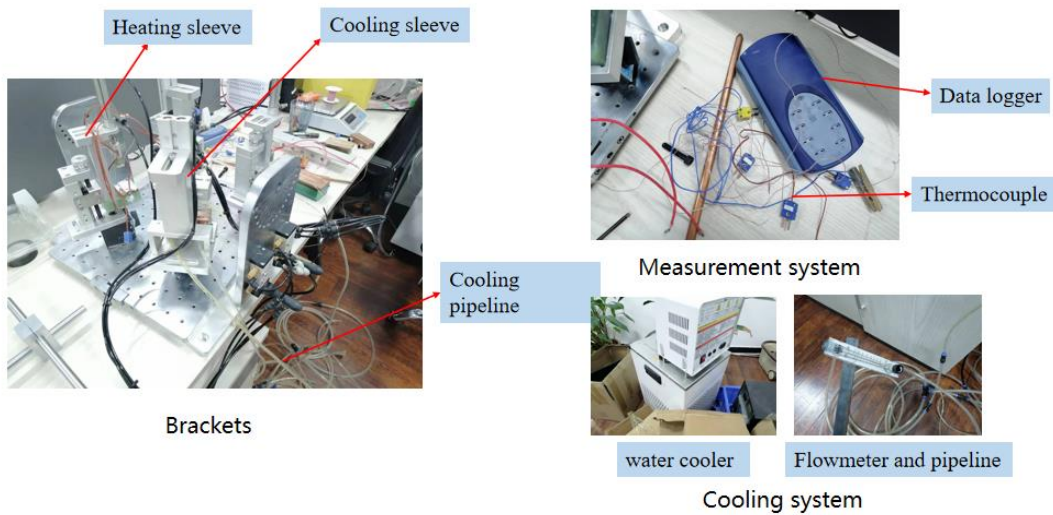

**Figure 5** Images of experimental system

Three types of heat pipes used in this experiment are 400 mm, 300 mm, and 200 mm in length, with an outer diameter of 8 mm for each heat pipe. The material of the heat pipe shell and capillary wick is copper, the wick structure is packed sphere, and the working fluid is water. The thermal conductivity coefficient of the 300 mm heat pipe is known as 2100 W/(m·K), while the characteristics of the other heat pipes are unknown.

In the experiment, the heat flow $Q$ inside the heat pipe is $Q = \eta P_\text{h}$, where $\eta$ is heating efficiency and $P_\text{h}$ is power of power supply. The equation for the temperature

difference between the evaporator and condenser sections of the heat pipe and the heating power is:

$$P_{\text{h}} = \frac{KA}{\eta l}(T_{\text{e}} - T_{\text{c}}) \tag{16}$$

where $K$ is thermal conductivity coefficient ($K = lk_{\text{HP}}$), $A$ is cross sectional area of heat pipe.

## 3. Results and Discussion

### 3.1 Analysis on the Calculated Heat Transfer Limits

It is difficult to see the influence of heat pipe characteristics on its heat transfer limit only based on the calculation equations. However, by using the heat transfer limit curve drawn by the program and changing a certain input characteristic through the single variable method, the impact of this characteristic on the heat pipe can be qualitatively analyzed.

The operating temperature range of a water heat pipe is approximately 0 to 300 °C. Figure 6 shows the heat transfer limit curve within the normal operating temperature range, and Table 2 is the characteristics of a wire screen wick water heat pipe.

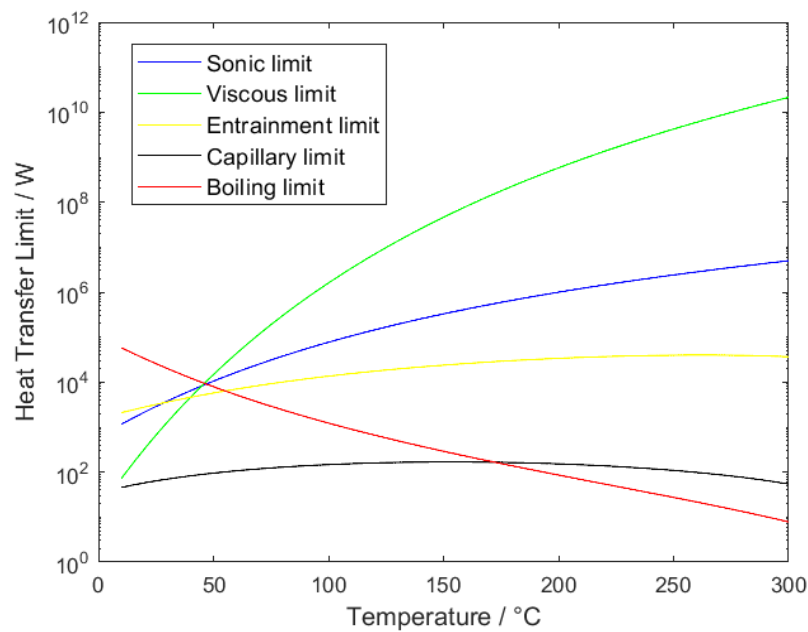

**Figure 6** Heat transfer limit curves of a water heat pipe

Table 2 Input parameters of the wire screen wick water heat pipe

| Characteristics | Input parameters |
| --- | --- |
| Inner diameter | $2.01 \times 10^{-2}$ m |
| Inclination angle | 0 ° |
| Condenser section length | 0.1 m |
| Adiabatic section length | 0.3 m |
| Evaporator section length | 0.1 m |
| Wick thickness | $1 \times 10^{-3}$ m |
| Wire diameter | $6.25 \times 10^{-5}$ m |
| Mesh count | 7870 |

The length of the primary heat pipe is 0.5 m. Figure 7 shows the total heat transfer limit curves of the heat pipes with a length of 0.25 m, 0.5 m, 0.75 m, and 1.0 m, by only changing the length of the heat pipe and keeping the length ratio of the evaporator section, adiabatic section, and condenser section unchanged.

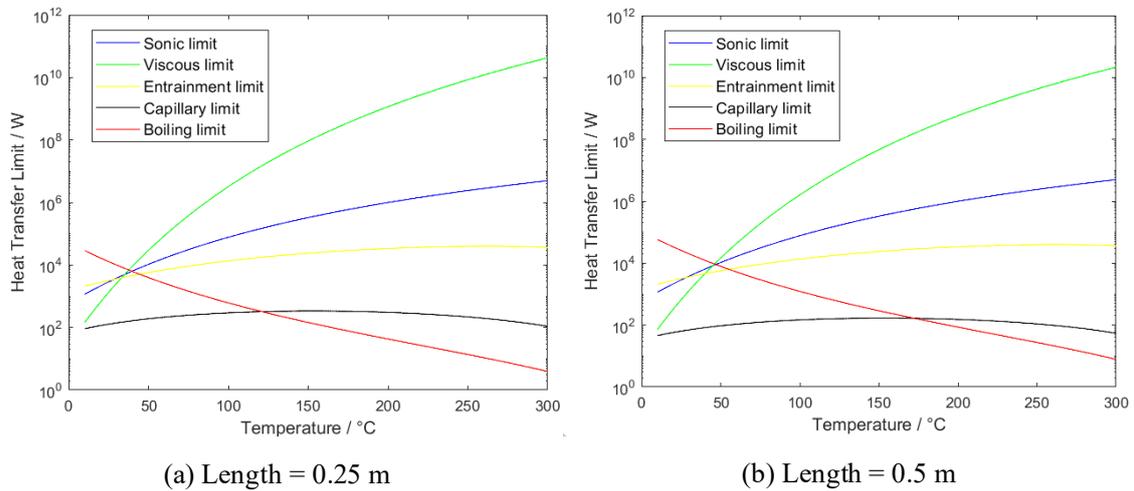

(a) Length = 0.25 m     (b) Length = 0.5 m

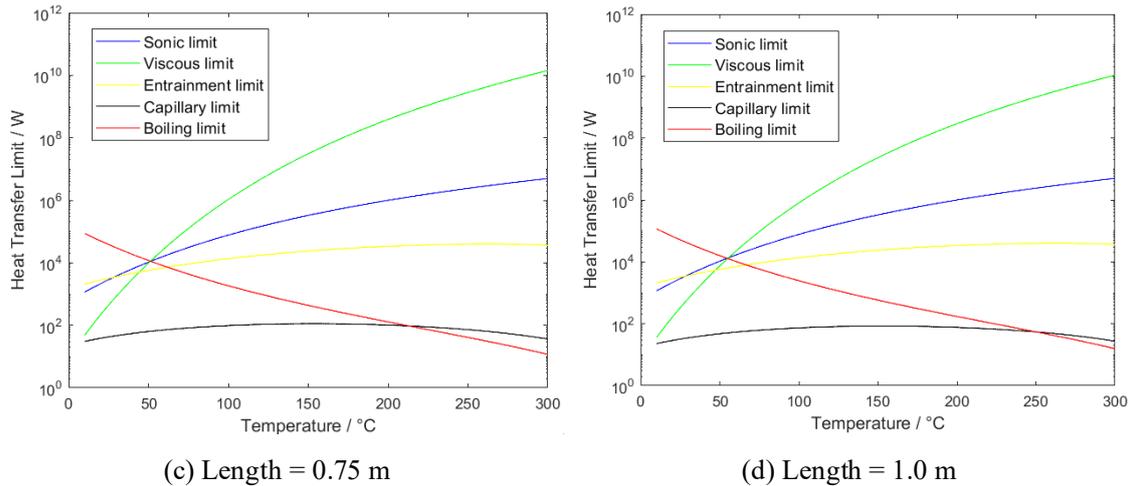

(c) Length = 0.75 m      (d) Length = 1.0 m

**Figure 7** Heat transfer limit of heat pipes with different lengths

From the above figures, it can be seen that the change in pipe length has nearly no effects on the sonic limit and entrainment limit. As the length increases, the viscous limit and capillary limit gradually decrease, while the boiling limit increases. An increase in the length of the heat pipe leads to an increase in the viscous and frictional forces in the vapor channel, resulting in a decrease in the viscous and capillary limit heat transfer. The increase in the length of the evaporator section leads to an increase in the total quantity of heat while maintaining the heat flux for boiling, thereby enhancing the heat flow at the boiling limit.

It is interesting that in the lower temperature range, it is the capillary limit that mainly restricts the heat transfer of the heat pipe, while in the high temperature range, the boiling limit plays a decisive role. As the length increases, the intersection point of the curve of boiling limit and capillary limit gradually shifts backwards. Therefore, when the heat pipe operates in the low-temperature range, increasing the pipe length reduces the overall heat transfer limit of the heat pipe, while increasing the pipe length in the high-temperature range can increase the heat transfer limit.

The heat transfer limit curves of the heat pipe within the normal operating temperature range for each wire diameter are obtained by taking wires with diameters of $5.00 \times 10^{-5}$ m, $6.25 \times 10^{-5}$ m, $7.50 \times 10^{-5}$ m and $8.75 \times 10^{-5}$ m, as shown in Figure 8.

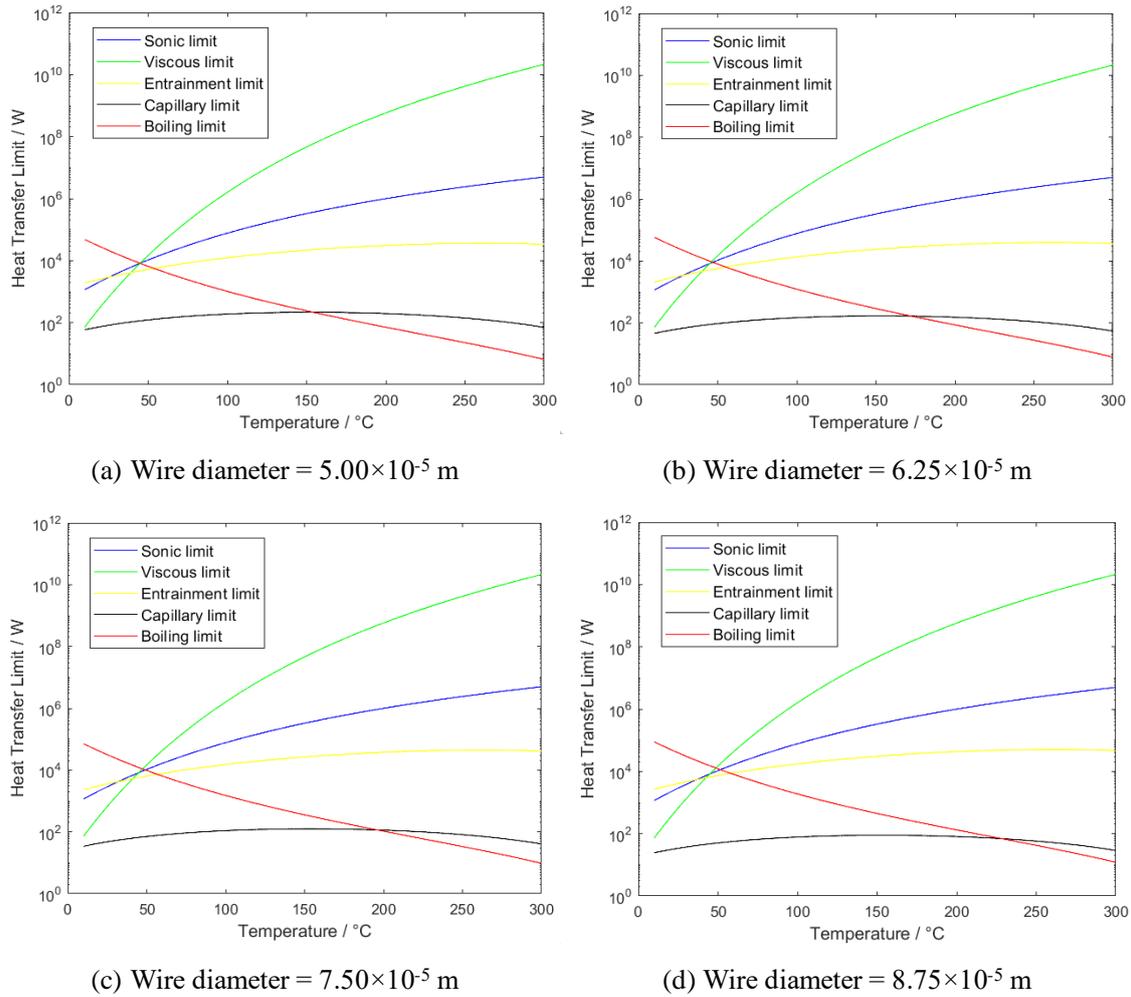

(a) Wire diameter = 5.00×10<sup>-5</sup> m  (b) Wire diameter = 6.25×10<sup>-5</sup> m

(c) Wire diameter = 7.50×10<sup>-5</sup> m  (d) Wire diameter = 8.75×10<sup>-5</sup> m

**Figure 8** Heat transfer limit of heat pipes with different wire diameters

From the figure above, it can be seen that as the wire diameter increases, the sonic limit, viscous limit, and entrainment limit remain unchanged, while the capillary limit gradually decreases and the boiling limit increases. The reason is likely that as the wire is thicker, the resistance of the liquid flow in the wick increases, leading to a decrease in capillary limit. However, due to the higher thermal conductivity of the copper wire compared to the working fluid, the overall thermal conductivity of the wick increases, resulting in a decrease in the temperature of the liquid and an increase in boiling limit. Overall, when the wire screen wick water heat pipe operates in the low temperature range, increasing the diameter of the wire will reduce the overall heat transfer limit of the heat pipe, while increasing the diameter of the in the high temperature range can increase the heat transfer limit of the heat pipe.

By changing the parameter of the inner diameter into $1.01\times10^{-2}$ m, $1.51\times10^{-2}$ m, $2.01\times10^{-2}$ m and $2.51\times10^{-2}$ m, the variation by heat transfer limits of different inner diameters can be observed in Figure 9. It seems that all the heat transfer limits increase as the inner diameter grows, mainly because of the increase of the vapor cross section area. One thing worth noticing is that the viscous limit might play a main role in limiting the heat transfer when heat pipe with small inner diameter working in low temperature, which is without precedent in Figure 7 and Figure 8.

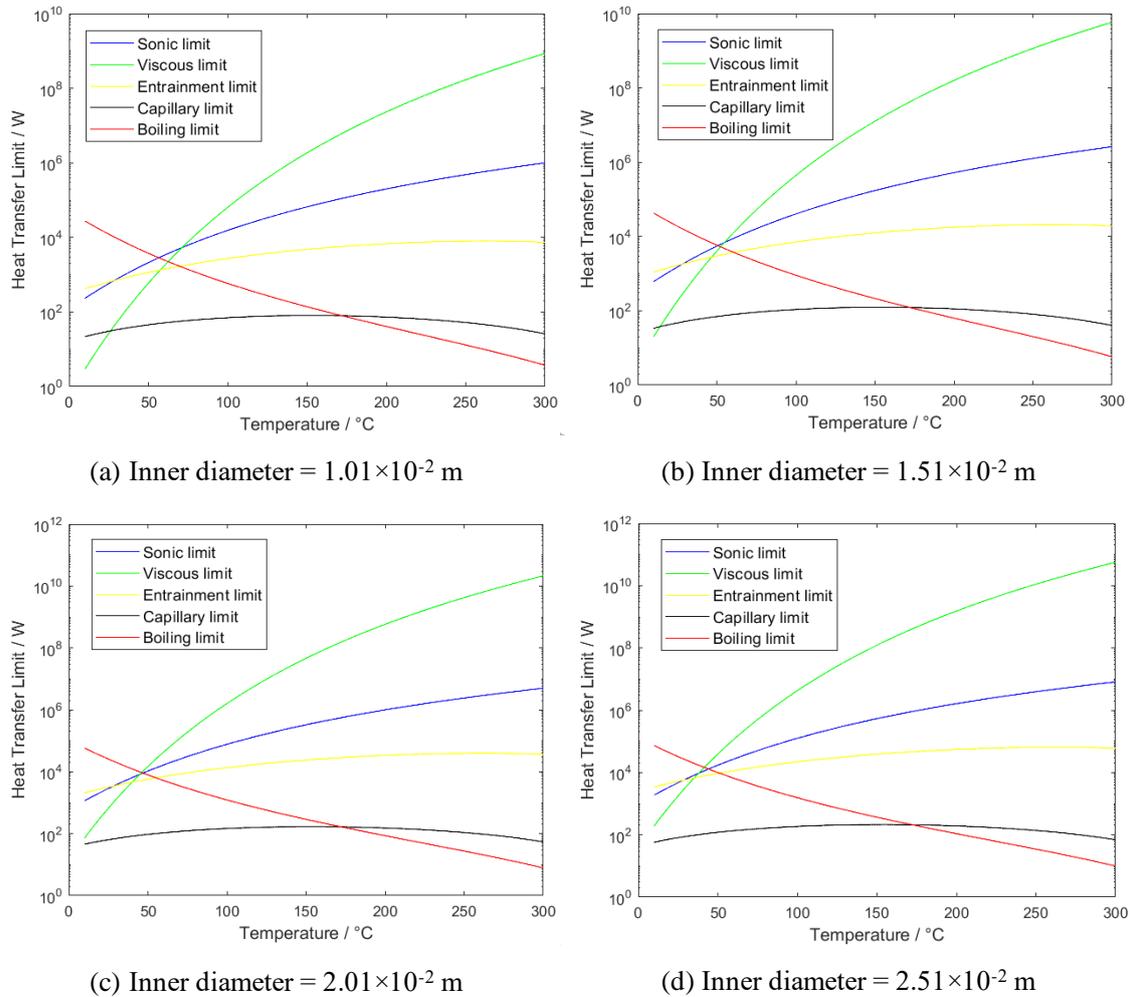

(a) Inner diameter = $1.01 \times 10^{-2}$ m  (b) Inner diameter = $1.51 \times 10^{-2}$ m

(c) Inner diameter = $2.01 \times 10^{-2}$ m  (d) Inner diameter = $2.51 \times 10^{-2}$ m

**Figure 9** Heat transfer limit of heat pipes with different inner diameter

In summary, except for the capillary limit and boiling limit, the other limits do not vary significantly with the changing of characteristics of the wick. The sonic limit and viscous limit are only related to the characteristics of vapor flow, and therefore are mainly affected by the physical properties of the working fluid and the size of the heat pipe; the entrainment limit is the result of the interaction between the surface tension of the liquid and the shear force at the gas-liquid interface, and is mainly determined by the properties of the working fluid, and it may also be related to the hydraulic radius of the suction core. In addition, the sonic limit is not affected by changes in length of heat pipes, because the critical section at which the vapor flow velocity reaches sonic speed is at the end of the evaporator section, which is not related to the heat pipe length but to the radius of the vapor channel.

### 3.2 Measurements of thermal conductance

Using the 300 mm heat pipe with known thermal conductivity for measurement, a regression curve between temperature difference and power can be plotted, and the slope of the curve can be calculated to obtain heating efficiency. The experiment measuring $\eta$ was conducted three times, and the data is shown in Figure 10.

The slope of the curve obtained from three experiments and the calculated heating efficiency $\eta$ are listed in Table 3. The average heating efficiency of the heating plate can be obtained from the table as 6.10%.

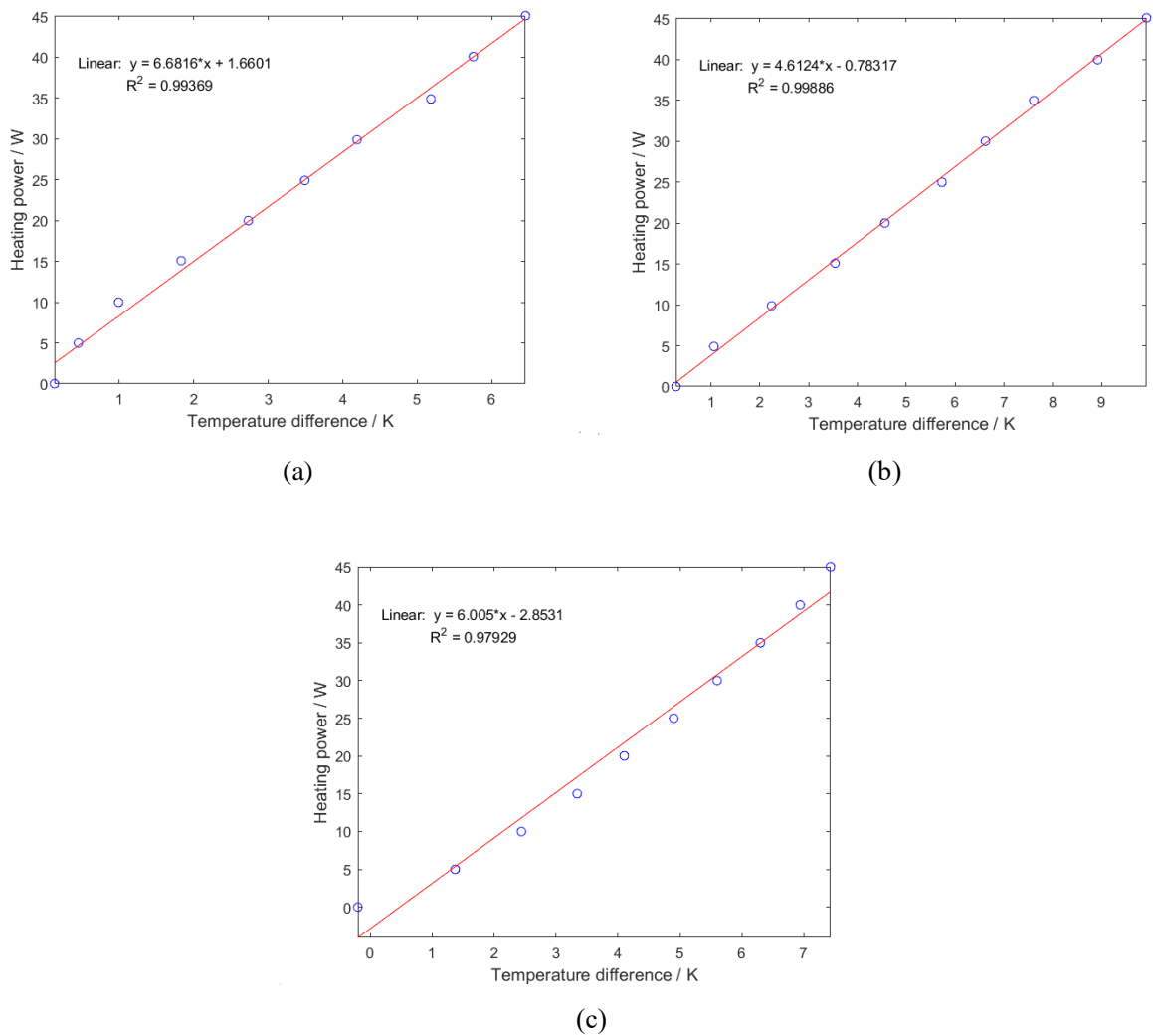

**Figure 10** Regression curve of heating power and temperature difference

Table 3 Heating efficiency of experimental system

| Measurement | Slope of regression curve | Heating efficiency $\eta$ |
|:---:|:---:|:---:|
| (a) | 6.6816 W/K | 5.27% |
| (b) | 4.6124 W/K | 7.63% |
| (c) | 6.0050 W/K | 5.86% |
| Average | 5.7663 W/K | 6.10% |

For the heat pipe with a known thermal conductivity of 2100 W/(m·K), its thermal conductance is 7000 W/(m²·K). For the other two heat pipes, their thermal conductivity and thermal conductance can be obtained from the slope of the regression curve between temperature difference and power.

Four experiments were conducted on the 400 mm heat pipe, and the third experiment collected incorrect experimental data due to the incorrect placement of the heat pipe on the heating plate. The data should be removed during data processing. Two experiments were conducted on a 200 mm heat pipe, and during the measurement process, it was found that its thermal conductivity was much lower than the other two heat pipes. It was suspected that this one may have been damaged, so no further experimental measurements were conducted. The slope, thermal conductivity coefficient, and thermal conductance obtained from experimental data are shown in Table 4.

Table 4 Experimental data, thermal conductivity and thermal conductance

(a) 400 mm heat pipe

| Measurement | Slope | Thermal conductivity | thermal conductance |
|:---:|:---:|:---:|:---:|
| (a) | 3.6308 W/K | 1762 W/(m·K) | 4405 W/(m²·K) |
| (b) | 6.6303 W/K | 3218 W/(m·K) | 8045 W/(m²·K) |
| (c) (invalid) | 13.268 W/K | / | / |
| (d) | 3.5151 W/K | 1706 W/(m·K) | 4265 W/(m²·K) |
| Average | 4.5921 W/K | 2229 W/(m·K) | 5572 W/(m²·K) |

(b) 200 mm heat pipe

| Measurement | Slope | Thermal conductivity | thermal conductance |
|---|---|---|---|
| (a) | 2.1614W/K | 525 W/(m·K) | 2625 W/(m²·K) |
| (b) | 2.8656W/K | 695 W/(m·K) | 3475 W/(m²·K) |
| Average | 2.5135W/K | 610 W/(m·K) | 3050 W/(m²·K) |

From the above table, it can be seen that the experimental data for the thermal conductivity of the 400 mm heat pipe is 2229 W/(m·K), which is close to that of the 300 mm heat pipe. This is because the processing technology for both is the same. The thermal conductivity of the 200mm heat pipe is 610 W/(m·K), while the thermal conductivity of copper is 413 W/(m·K). Its heat transfer ability is similar to that of ordinary thermal conductive materials, so it can be concluded that this heat pipe has been damaged.

If the porosity of the packed sphere wick is 0.524 (obtained based on the close packing of identical spheres), then according to the previous equation for the thermal conductivity of capillary wick, 2.45 W/(m·K) can be obtained. According to equation 15, the theoretical thermal conductance for 400 mm is approximately 7267 W/(m²·K). There is a deviation from the experimentally measured data of 5572 W/(m²·K), which is mainly due to inaccurate estimation of some characteristics, but this deviation is within an acceptable range.

## 4. Conclusion

In an effort to describe the heat transfer characteristics of heat pipe, its heat transfer limits and thermal conductance were investigated by both theoretical and experimental methods.

By calculating the heat transfer performance of heat pipes with different characteristics, it is found that the changes in the characteristics of the capillary wick mainly affect the capillary limit and boiling limit, while the sonic limit and viscous limit are only related to the characteristics of vapor flow. As the length increases, the viscous limit and capillary limit gradually decrease, while the boiling limit increases. For the wire screen wick heat pipe, as the diameter of the wire increases, the capillary limit gradually decreases and the boiling limit increases. According to the heat transfer pathway, a one-dimensional thermal resistance model was established, which was used for deriving the equation of heat pipe thermal conductance. An experimental setup which could measure the thermal conductance and heat transfer limit of heat pipes was conducted to validate the theoretical models. And by comparing the theoretical result of the equation with the experimental result, the heat pipe thermal conductance equation was proved to be effective within an acceptable error range.

The heat transfer models presented in this work can be applied in designing heat pipe to know its heat transfer characteristics. And the experimental setup can be improved for further measurement such as the curve of heat transfer limit of heat pipe.

# Appendix

In this section, the derivations of different models of heat transfer limits are briefly reviewed.

### A1. Sonic Limit and Viscous Limit

Considering the vapor inside the heat pipe as the ideal gas, the equation of state is:

$$\frac{P}{\rho} = \frac{P_0}{\rho_0} = \frac{RT_0}{M} \tag{A1-1}$$

and the axial heat flow can be described by:

$$q = \rho_v \overline{w} \lambda \tag{A1-2}$$

in which $\overline{w}$ is the mean axial vapor flow velocity on the cross-section.

For the axial pressure drop the following equations hold[4]: in the inertia flow regime:

$$\Delta P(x) = \rho(x)\overline{w^2(x)} \tag{A1-3}$$

in the viscous flow regime:

$$\Delta P(x) = -\frac{32\mu_v}{d^2}\overline{w}(x) \tag{A1-4}$$

which are derived from Navier-Stokes equation.

Combining Equation (17), (18) and (19) it follows that:

$$q = \frac{\lambda(\rho_0 P_0)^{\frac{1}{2}}}{\bar{A}^{\frac{1}{2}}}\left[\frac{P}{P_0}\left(1 - \frac{P}{P_0}\right)\right]^{\frac{1}{2}} \tag{A1-5}$$

in which $\bar{A}$ is the ratio of vapor mean square velocity to the square of average velocity. It can be seen only that $q$ evidently has a maximum for a non-vanishing value of $P$, that choking of the vapour flow occurs. The condition for this maximum, which is the sonic limit of heat transfer, is obtained by putting $dq/dP = 0$, so it follows from (21) that the average axial heat flux at the sonic limit of heat transfer which is shown in Eq. (1)

Combining Eqs. (17), (18), (20) and integrating along the heat pipe it follows that:

$$q = \frac{d^2\lambda}{64\mu_v l_{eff}}(1 - \frac{P^2}{P_0^2})\rho_0 P_0 \tag{A1-6}$$

the heat pipe reaches the viscous limit when $P/P_0 = 0$. Then Eq. (2) is obtained.

## A2. Capillary Limit

The equilibrium equation between maximum capillary pressure and working fluid transport pressure in heat pipe is[5]:

$$\frac{2\sigma}{r_c} - \rho_l g d \cos\varphi = \int_0^l (\frac{dP_v}{dx} - \frac{dP_l}{dx})dx \tag{A2-1}$$

Utilizing heat pipe vapor flow model and liquid flow model[8], Equation (23) can be rewritten as:

$$\frac{2\sigma}{r_c} - \rho_l g d \cos\varphi = \int_0^l (F_v Q + F_l Q + \rho_l g \sin\varphi)dx \tag{A2-2}$$

in which $F_v$ and $F_l$ are both friction coefficient of vapor and fluid. Then the capillary limit of heat transfer is now obtained as shown in Eq. (4).

## A3. Entrainment Limit

The shear force at the liquid–vapor interface $F_s$ and tends to tear the liquid from the surface of the wick is proportional to the product of dynamic pressure of the moving vapor $\rho w^2/2$ and the area $A_s$ of individual surface pores of the wick[5]:

$$F_s = K_1 \frac{\rho w^2 A_s}{2} \tag{A3-1}$$

and the surface force $F_t$ which holds the liquid in the wick is:

$$F_t = K_2 C_s \sigma \tag{A3-2}$$

in which $C_s$ is the wetted perimeter.

In both Equation (25) and (26), the constant proportionality is given by $K_1$ and $K_2$ respectively. The conditions leading to entrainment are expressed in terms of the ratio of vapor inertial forces to liquid surface tension forces, called the Weber number.

When the Weber number reaches unity that is given by:

$$We = \frac{K_1 \rho w^2 A_s}{2 K_2 C_s \sigma} = 1 \tag{A3-3}$$

Experimental data show the value of $K_1/K_2$ is of the order of 8[8]. And the hydraulic radius $r_h$ is twice the $A_s$ divided by the $C_s$, then entrainment limit can be written as:

$$We = \frac{2 r_h \rho w^2}{\sigma} = 1 \tag{A3-4}$$

The vapor velocity $w$ of a heat pipe is related to the axial heat flux by Equation (18), then the expression for the entrainment heat transport limit is as shown in Eq. (5).

**A4. Boiling Limit**

Assume that a vapor bubble sphere at vicinity of wick structure rises while it is interfacing the wick and it is in equilibrium state, then the following equation is a valid expression:

$$\pi r_b^2 (P_s - P_l) = 2\pi r_b \sigma \tag{A4-1}$$

where $P_s$ is saturation vapor pressure at heat pipe wick interface, $P_l$ is liquid pressure which is equal to vapor pressure minus capillary pressure $(P_v - P_c)$. Utilizing the Clausius–Clapeyron equation relates T and P along the saturation line as follows:

$$\frac{dP}{dT} = \frac{\lambda \rho}{T_v} \tag{A4-2}$$

Assuming that $P_s - P_v \approx \Delta T dP/dT$, now it gets the result as:

$$\Delta T = \frac{T_v}{\lambda \rho} \left( \frac{2\sigma}{r_b} - P_c \right) \tag{A4-3}$$

where $\Delta T$ is the temperature drop across the wick structure at the evaporator section. Utilize conduction theorem for a heat pipe with uniform heat flux distribution along the evaporator length $l_e$:

$$\Delta T = \frac{Q \ln(r_i/r_v)}{2\pi l_e k_\varepsilon} \tag{A4-4}$$

to get the following result:

$$Q = \frac{2\pi l_e k_\varepsilon T_v}{\lambda \rho \ln(r_i/r_v)} \left( \frac{2\sigma}{r_b} - P_c \right) \tag{A4-5}$$

where $P_c$ is negligible compared to $2\sigma/r_b$, then the boiling limit is as shown in Eq. (6).

# Nomenclature

| | | | |
|---|---|---|---|
| $A$ | cross sectional area of heat pipe | $r_b$ | nucleation site radius |
| $A_v$ | cross-sectional area of vapor flow channel | $r_c$ | effective capillary radius |
| $d$ | inner diameter of heat pipe | $r_h$ | hydraulic radius of the wick surface pores |
| $F_l$ | liquid frictional coefficient | $r_i$ | inner radius of heat pipe |
| $F_v$ | vapor frictional coefficient | $r_o$ | outer radius of heat pipe |
| $k_l$ | thermal conductivity coefficient of liquid working fluid | $r_v$ | radius of vapor flow channel |
| $k_p$ | thermal conductivity coefficient of pipe shell | $t_p$ | thickness of pipe shell |
| $k_w$ | thermal conductivity coefficient of wick material | $t_w$ | thickness of wick |
| $k_{HP}$ | thermal conductance of heat pipe | $T_c$ | vapor temperature of condenser section |
| $k_\varepsilon$ | thermal conductivity coefficient of liquid-saturated wicks | $T_e$ | vapor temperature of evaporator section |
| $K$ | thermal conductivity coefficient of heat pipe | $T_v$ | vapor temperature |
| $l$ | length of heat pipe | $\delta$ | groove depth |
| $l_a$ | length of adiabatic section | $\varepsilon$ | porosity |
| $l_c$ | length of condenser section | $\eta$ | heating efficiency |
| $l_e$ | length of evaporator section | $\lambda$ | latent heat of vaporization |
| $l_{eff}$ | effective length of heat pipe | $\mu_v$ | vapor viscosity |
| $P_0$ | vapor pressure at the evaporator exit | $\rho_0$ | vapor density at the evaporator exit |
| $P_t$ | maximum capillary pressure | $\rho_l$ | liquid density |
| $P_{v,c}$ | vapor pressure at condenser section | $\rho_v$ | vapor density |
| $P_{v,e}$ | vapor pressure at evaporator section | $\sigma$ | surface tension coefficient |
| $P_h$ | power of power supply | $\varphi$ | heat pipe inclination |

*Q*    heat flow or heat transfer limit

# Acknowledgement

This work was supported by the National Natural Science Foundation of China [No.12375170] and Natural Science Foundation of Beijing [No. 1242020]. H.Z. is supported by ZHONGHEYINGCAI of 2023; State Key Laboratory of Chemical Engineering (No. SKL-ChE-23A03) and Tsinghua University Initiative Scientific Research Program.